\documentclass[12pt,epsf]{article}
\usepackage{verbatim}

\usepackage{dsfont}
\usepackage{sidecap}
\sidecaptionvpos{figure}{t}
\usepackage{subfigure}

\usepackage[english]{babel}
\usepackage{enumitem}  

\usepackage{amssymb,amsmath}
\usepackage{graphicx, xcolor, varwidth}
\usepackage{setspace}
\usepackage[permil]{overpic}
\usepackage{cite}

\usepackage{feynmp}
\DeclareSymbolFont{matha}{OML}{txmi}{m}{it}
\DeclareMathSymbol{v}{\mathord}{matha}{118}

\colorlet{darkblue}{blue!70!black}
\colorlet{darkgreen}{green!70!black}

\usepackage[colorlinks=true,urlcolor=darkblue,linktocpage=true,linkcolor=darkblue,citecolor=darkblue]{hyperref}

\numberwithin{equation}{section}

\newcommand{\be}{\begin{equation}}
\newcommand{\ee}{\end{equation}}
\newcommand{\bea}{\begin{eqnarray}}
\newcommand{\eea}{\end{eqnarray}}
\newcommand{\bear}{\begin{eqnarray}}
\newcommand{\eear}{\end{eqnarray}}  
\newcommand{\beas}{\begin{eqnarray*}}
\newcommand{\p}{\partial}
\newcommand{\eeas}{\end{eqnarray*}}
\newcommand{\ba}{\begin{array}}
\newcommand{\ea}{\end{array}}

\newcommand{\nn}{\nonumber}

\newcommand{\pd}[2][1]{\ifnum#1=1 \frac{\partial}{\partial {#2}} \else
  \frac{\partial^#1}{\partial {#2}^{#1}}\fi}
\newcommand{\dpd}[2][1]{\ifnum#1=1 \dfrac{\partial}{\partial {#2}} \else
  \frac{\partial^#1}{\partial {#2}^{#1}}\fi}
\newcommand{\td}[2][1]{\ifnum#1=1 \frac{d}{d{#2}} \else
  \frac{d^#1}{d{#2}^{#1}}\fi}

\newcommand{\smallstart}{
\end{spacing}
\noindent\hfil\rule{1\textwidth}{.4pt}\hfil\small

   \addtolength{\leftskip}{5mm}
}
\newcommand{\smallend}{
   \addtolength{\leftskip}{-5mm}
\noindent\hfil\rule{1\textwidth}{.4pt}\hfil\normalsize
\begin{spacing}{1.3}
}

\newcommand{\nbox}{{\,\lower0.9pt\vbox{\hrule \hbox{\vrule height 0.2 cm \hskip 0.19 cm \vrule height 0.2 cm}\hrule}\,}}

\newcommand{\ie}{{\it i.e.,}\ }
\newcommand{\E}{{\bf E}}

\newcommand{\bTheta}{\mathbf{\Theta}}

\newcommand{\bz}{\bar{z}}

\newcommand{\bO}{\mathcal{O}}

\renewcommand{\Im}{\mbox{Im\ }}
\renewcommand{\Re}{\mbox{Re\ }}
\renewcommand{\bTheta}{\overline{\Theta}}



\newcommand{\sdot}{\!\cdot\!}

\renewcommand{\E}{\mathcal{E}}
\newcommand{\etal}{\textit{et al}.}

\newcommand{\e}{\varepsilon}

\begin{document}
\begin{spacing}{1.3}
\begin{titlepage}

\begin{center}
{\Large \bf Averaged Null Energy Condition from Causality}

\vspace*{12mm}

Thomas Hartman, Sandipan Kundu, and Amirhossein Tajdini \\

\vspace*{6mm}

\textit{Department of Physics, Cornell University, Ithaca, New York\\}

\vspace{6mm}

{\tt hartman@cornell.edu, kundu@cornell.edu, at734@cornell.edu}

\vspace*{15mm}
\end{center}
\begin{abstract}

 Unitary, Lorentz-invariant quantum field theories in flat spacetime obey microcausality: commutators vanish at spacelike separation. For interacting theories in more than two dimensions, we show that this implies that the averaged null energy, $\int du T_{uu}$, must be positive. This non-local operator appears in the operator product expansion of local operators in the lightcone limit, and therefore contributes to $n$-point functions.  We derive a sum rule that isolates this contribution and is manifestly positive. The argument also applies to certain higher spin operators other than the stress tensor, generating an infinite family of new constraints of the form $\int du X_{uuu\cdots u} \geq 0$. These lead to new inequalities for the coupling constants of spinning operators in conformal field theory, which include as special cases (but are generally stronger than) the existing constraints from the lightcone bootstrap, deep inelastic scattering, conformal collider methods, and relative entropy. We also comment on the relation to the recent derivation of the averaged null energy condition from relative entropy, and suggest a more general connection between causality and information-theoretic inequalities in QFT.

\end{abstract}

\end{titlepage}
\end{spacing}

\vskip 1cm

\setcounter{tocdepth}{2}
\tableofcontents

\begin{spacing}{1.3}
\newpage

\section{Introduction}

The average null energy condition (ANEC) states that 
\be
\int_{-\infty}^\infty d\lambda\,  \langle T_{\alpha\beta}\rangle u^\alpha u^\beta \geq 0 \ ,
\ee
where the integral is over a complete null geodesic, and $u$ is the tangent null vector. This inequality plays a central role in many of the classic theorems of general relativity \cite{Borde:1987qr,Roman:1986tp,Frolov:1998wf}. Matter violating the ANEC, if it existed, could be used to build time machines \cite{Morris:1988tu,Friedman:1993ty} and violate the second law of thermodynamics \cite{Wall:2009wi}. And, unlike most of the other energy conditions discussed in relativity (dominant, strong, weak, null, etc.), the ANEC has no known counterexamples in consistent quantum field theories (assuming also that the null geodesic is achronal \cite{Graham:2007va}).

Though often discussed in the gravitational setting, the ANEC is a statement about QFT that is nontrivial even in Minkowski spacetime without gravity. In this context, the first general argument for the ANEC in QFT was found just recently by Faulkner, Leigh, Parrikar, and Wang \cite{Faulkner:2016mzt}. (Earlier derivations \cite{Klinkhammer:1991ki,Wald:1991xn,Folacci:1992xg,Verch:1999nt,Bousso:2015wca}, were restricted to free or superrenormalizable theories, or to two dimensions.) The crucial tool in the derivation of Faulkner \etal\ is monotonicity of relative entropy. Assuming all of the relevant quantities are well defined in the continuum limit, the argument applies to a large (and perhaps dense) set of states in any unitary, Lorentz-invariant QFT.
  
Separately, the ANEC for a special class of states in conformal field theory was derived recently using techniques from the conformal bootstrap developed in \cite{mss,causality1}. These special cases of the ANEC, known as the Hofman-Maldacena conformal collider bounds \cite{Hofman:2008ar}, were derived in \cite{causality2,Hofman:2016awc}.  The derivation relied on causality of the CFT, in the microscopic sense that commutators must vanish outside the lightcone, applied to the 4-point correlator $\langle \phi[T,T]\phi\rangle$ where $T$ is the stress tensor and $\phi$ is a scalar. However, it was not clear from the derivation why the bootstrap agreed with the ANEC as applied by Hofman and Maldacena, or whether there was a more general connection between causality and the ANEC in QFT.

In this paper, we simplify and extend the causality argument and show that it implies the ANEC more generally. We conclude that any unitary, Lorentz-invariant QFT with an interacting conformal fixed point in the UV must obey the ANEC, in agreement with the information-theoretic derivation of Faulkner \etal\ The argument assumes no higher spin symmetries at the UV fixed point, so it requires $d>2$ spacetime dimensions and does not immediately apply to free (or asymptotically free) theories. A byproduct of the analysis is a sum rule for the integrated null energy in terms of a manifestly positive 4-point function. Furthermore, we argue that the ANEC is just one of an infinite class of positivity constraints of the form
\be\label{anoc}
\int du X_{uu \cdots u} \geq 0
\ee
where $X$ is an even-spin operator on the leading Regge trajectory (normalized appropriately) --- \textit{i.e.,} it is the lowest-dimension operator of spin $s \geq 2$. 
This implies new constraints on 3-point couplings in CFT; we work out the example of spin-1/spin-1/spin-4 couplings. Another interesting corollary is that, like the stress tensor, the minimal-dimension operator of each even spin must couple with the same sign to all other operators in the theory. (There may be exceptions under certain conditions; see section \ref{GSO} for a discussion of the subtleties.) In analogy with the Hofman-Maldacena conditions on stress tensor couplings, we conjecture that \eqref{anoc} evaluated in a momentum basis is optimal, meaning that the resulting constraints on 3-point couplings can be saturated in consistent theories. This remains to be proven.\footnote{In free field theory our methods do not apply directly, but a simple mode calculation in an appendix demonstrates that the inequality \eqref{anoc} holds also for free scalars. This appears to have escaped notice. It may have interesting implications for coupling quantum fields to stringy background geometry, just as the ANEC plays an important role in constraining physical spacetime backgrounds.  The operators $X$ generalize the stress tensor to the full leading Regge trajectory of the closed string. A first step would be to confirm that \eqref{anoc} holds for other types of free fields.}

The connection between causality and null energy is well known in the gravitational context (see for example \cite{Gao:2000ga,Dubovsky:2005xd} and the references above) and in AdS/CFT (see for example \cite{Kelly:2014mra,Engelhardt:2016aoo}). In a gravitational theory, null energy can backreact on the geometry in a way that leads to superluminal propagation in a curved background.  Our approach is quite different, since we work entirely in quantum field theory, without gravity, and invoke microcausality rather than superluminal propagation in curved spacetime. 

\subsubsection*{Causality vs. Quantum Information}
Our derivation bears no obvious resemblance to the relative entropy derivation of Faulkner \etal, except that both seem to rely on Lorentzian signature. (Our starting point is Euclidean, and we do not make any assumptions about the QFT beyond the usual Euclidean axioms, but we do analytically continue to Lorentzian.) It is intriguing that causality and information-theoretic inequalities lead to overlapping constraints in this context. 

There are significant hints that this connection between entanglement and causality is more general. This is certainly true in 2d CFT; see for example \cite{Cardy:2015xaa}. It is also clear in general relativity; for example, Wall showed that both the second law \cite{Wall:2009wi} and strong subadditivity of the holographic entanglement entropy \cite{Wall:2012uf} require the ANEC.  But there are also hints in higher-dimensional QFT for a deeper connection between entanglement and causality constraints. Recent work on the quantum null energy condition \cite{Bousso:2015mna,Bousso:2015wca} is one example, and they are also linked by $c$-theorems for renormalization group flows in various dimensions. The $F$-theorem, which governs the renormalization group in three dimensions \cite{Myers:2010tj,Jafferis:2011zi,Casini:2012ei}, was derived from strong subadditivity of entanglement entropy but has resisted any attempt at a derivation using more traditional tools.  On the other hand, its higher-dimensional cousin, the $a$-theorem in four dimensions, was derived by  invoking a causality constraint \cite{Komargodski:2011vj} (and in this case, attempts to construct an entanglement proof have been unsuccessful). So causality and entanglement constraints both tie deeply to properties of the renormalization group, albeit in different spacetime dimensions. Another tantalizing hint is that in holographic theories, RG monotonicity theorems in general dimensions are equivalent to causality in the emergent radial direction \cite{Myers:2010tj}. 

These clues suggest that the two types of Lorentzian constraints --- from causality and from quantum information --- are two windows on the same phenomena in quantum field theory. It would be very interesting to explore this further. For instance, perhaps the $F$-theorem can be understood from causality; after all, a holographic violation of the $F$-theorem would very likely violate causality, too. It also suggests that the higher-spin causality constraints \eqref{anoc} on the leading Regge trajectory could have an information-theoretic origin, presumably involving non-geometric deformations of the operator algebra.

\subsubsection*{Comparison to previous methods}

Both conceptually and technically, the argument presented here has several advantages over previous bootstrap methods in \cite{causality1,causality2,Hofman:2016awc}. First, it makes manifest the connection between causality constraints and integrated null energy. Second, it produces optimal constraints (for example the full Hofman-Maldacena bounds on $\langle TTT\rangle$) without the need to decompose the correlator into a sum over composite operators in the dual channel.   This decomposition, accomplished in \cite{Hofman:2016awc}, was technically challenging for spinning probes, and becomes much more unweildy with increasing spins (say, for $\langle TTTT\rangle$). The simplification here comes from the fact that the new approach allows for smeared operator insertions, and these can be used to naturally project out an optimal set of positive quantities.
Finally, the new method produces stronger constraints on the 3-point functions of non-conserved spinning operators. On the other hand, this approach does not give us the solution of the crossing equation in the lightcone limit or the anomalous dimensions of high-spin composite operators as in \cite{Hofman:2016awc}.

\subsubsection*{Outline}
The main argument is given first, in section \ref{s:anec}. The essential new ingredient that it relies on is the fact that the null energy operator appears in the lightcone OPE; this is derived in section \ref{s:ope}. For readers already familiar with the chaos bound \cite{mss} and/or earlier causality constraints \cite{causality1}, sections \ref{s:anec} - \ref{s:ope} give a complete derivation of $\int du T_{uu} \geq 0$. The sum rule is derived in section \ref{s:properties}, where we also review the methods of \cite{mss,causality1}. In section \ref{s:hm}, we show how to smear operators to produce directly the conformal collider bounds in the new approach --- this section is in a sense superfluous because conformal collider bounds follow from the ANEC, but it is useful to see directly how the two methods compare. In all cases we are aware of, this particular smearing produces the optimal set of constraints on CFT 3-point couplings. Finally, in section \ref{GSO}, we generalize the argument to the ANEC in any dimension $d>2$, as well as to an infinite class of higher spin operators $X$.

\section{Derivation of the ANEC}\label{s:anec}

In this section we outline the main argument. Various intermediate steps are elaborated upon in later sections. Our conventions for points $x \in R^{1,d-1}$ are
\be
x = (t, y, \vec{x}) \quad \mbox{or} \quad (u,v,\vec{x}), \quad u=t-y , \quad v = t+y  \ .
\ee
In expressions where some arguments are dropped, those coordinates are set to zero. $\psi$ is always a real scalar primary operator. Hermitian conjugates written as $O^\dagger(x)$ act only on the operator, not the coordinates, so $[O(t,\dots)]^\dagger = O^\dagger(t^*,\dots)$. To simplify the formulas we set $d=4$ in the first few sections, leaving the general case ($d>2$) for section \ref{GSO}.

Define the average null energy operator
\be\label{defE}
\E = \int_{-\infty}^{\infty}du T_{uu}(u,v=0, \vec{x}=0) \ .
\ee
The goal is to show that this is positive in any state, $\langle \Psi| \E | \Psi\rangle \geq 0$. We first discuss conformal field theories.  In CFT, it is sufficient to show that
\be\label{opos}
\langle O^\dagger(t=i\delta,y=0)  \, \E \, O(t=-i\delta,y=0)\rangle \geq 0
\ee
for an arbitrary local operator $O$ (not necessarily primary).  The insertion of $O$ in imaginary time creates a state on the $t=0$ plane so that this 3-point function can be interpreted as an expectation value $\langle O| \E | O\rangle$. And, in a conformal field theory, a dense subset of normalizable states can be created this way.\footnote{On the sphere, these states are complete by the state-operator correspondence. On the plane, therefore, any state the consists of local operators smeared over some finite region can be created this way, and by the Reeh-Schlieder theorem, such states are dense in the Hilbert space \cite{haag}.}

\begin{figure}
\begin{center}
\includegraphics{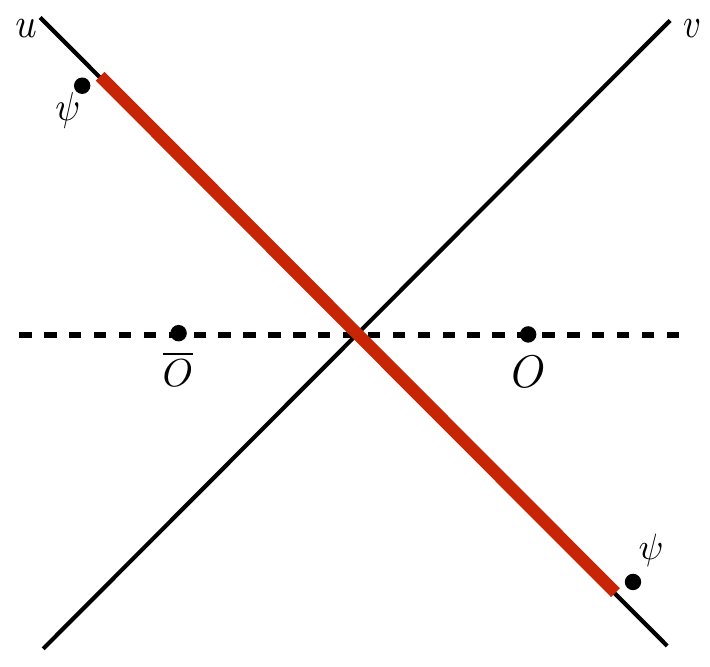}
\end{center}
\caption{\small Kinematics for the derivation of the ANEC. The leading correction to the $\psi\psi$ OPE is the null energy integrated over the red line, which in the limit of large $u$ takes the form $\langle  \overline{O}  \int du T_{uu} O\rangle$. This is then related to an expectation value by a Euclidean rotation.\label{fig:fourpoint}}
\end{figure}

In section \ref{s:ope}, we show that the non-local operator $\E$ makes a universal contribution to correlation functions in the lightcone limit.  The key observation is that the operator product expansion of two scalars in the lightcone limit can be recast as
\be\label{nulleope0}
\psi(u,v)\psi(-u,-v) \approx \langle \psi(u,v)\psi(-u,-v)\rangle  (1+ \lambda_T vu^2 \E ) \ ,
\ee
where $\lambda_T =\frac{10 \Delta_\psi}{c_T \pi^2} $ ($c_T$ is the coefficient of the $TT$ two-point function). This is the leading term in the regime
\be\label{uvlim}
 \qquad |v| \ll \frac{1}{|u|} \ll 1 \qquad (\mbox{with\ } uv < 0) \ .
\ee
The result \eqref{nulleope0} is an operator equation that can be used inside correlation functions (subtleties are discussed below). It sums the usual lightcone OPE (studied for example in \cite{Komargodski:2012ek,Fitzpatrick:2012yx}) including the contributions of all minimal-twist operators, $(\p_u)^n T_{uu}$ for $n \geq 0$. We have assumed that the theory is interacting, so there are no conserved currents of spin $> 2$ \cite{Maldacena:2011jn}, and that there are no very low-dimensions scalars in this OPE. (The second assumption is not necessary for the derivation of the sign constraints since we can project onto stress tensor exchange; see section \ref{GSO}.) 

Now consider the normalized 4-point function
\be\label{maing}
G = \frac{\langle \overline{O(y=\delta)} \psi(u, v)\psi( -u, - v) O(y=\delta)\rangle}{\langle \overline{O(y=\delta)}O(y=\delta)\rangle \langle  \psi(u, v)\psi( -u, - v) \rangle} \ ,
\ee
in the regime \eqref{uvlim}, as illustrated in figure \ref{fig:fourpoint}. $ \overline{O}$ denotes the Rindler reflection of the operator $O$. For scalars, $ \overline{O(t,y,\vec{x})} \equiv O^\dagger(-t, -y, \vec{x})$; see section \ref{RP} for the action on spinning operators. The OPE \eqref{nulleope0} gives
\be\label{gcor}
G = 1 + \frac{\lambda_T}{N_\delta} v u^2  \langle  \overline{O(y=\delta)} \, \E \, O(y=\delta) \rangle 
\ee
with $N_\delta = \langle  \overline{O(y=\delta)}O(y=\delta)\rangle > 0 $. The correction term, $\langle  \overline{O(y=\delta)} \, \E \, O(y=\delta) \rangle $, is computed by a residue of the null line integral, and is purely imaginary.

Although the correction in \eqref{gcor} is small, it is growing with $u$. Corrections of this form were studied by Maldacena, Shenker, and Stanford in \cite{mss}, and in the CFT context in \cite{causality1,causality2}, where it was shown that if such a term appears, it must have a negative imaginary part.\footnote{The chaos bound of MSS refers to large-$N$ theories, and the interest in \cite{mss} was in a different regime of the correlator (the Regge regime). Here, as in \cite{causality1}, we are applying similar methods to the lightcone regime of a small-$N$ theory. In the Regge/chaos regime, the small parameter that controls the OPE is $1/N$, whereas here it is the expansion in $v$ as we approach the lightcone. These limits do not, in general, commute, even in large $N$ theories, so the physics of the two classes of constraints is different.}  Therefore
\be\label{anecoo}
i \langle  \overline{O(y=\delta)} \, \E \, O(y=\delta) \rangle \ge 0\ .
\ee
This conclusion relies on a number of analyticity and positivity conditions that the correlator must satisfy; we check these conditions and review the argument in detail below. It can be understood as a causality constraint. If the correction has the wrong sign, then it requires the correlator to have singularities in a disallowed regime, and these singularities lead to non-vanishing commutators at spacelike separation. 

This is not yet \eqref{opos}, since in one case the operators are inserted in Minkowski space and in the other case offset in imaginary time. In fact, these are equivalent, by acting with a rotation $R$ that rotates by $\frac{\pi}{2}$ in the Euclidean $y\tau$-plane (with $\tau = i t$): 
\be\label{orotate}
i\langle \overline{O(y=\delta)} \, \E \, O(y=\delta) \rangle= \langle (R\sdot O)^{\dagger}(t=i\delta) \ \E\  R\sdot O(t=-i\delta)\rangle  \ .
\ee
(The null contour defining $\E$ is also trivially rotated in relating these two expressions.) The ANEC, in an arbitrary state in CFT, then immediately follows from \eqref{anecoo}.  

For comparison to previous work, we note that the arguments in \cite{mss,causality1,causality2} were phrased in terms of conformal cross ratios, and it was important that the correlator was evaluated on the `2nd sheet', \ie after a particular analytic continuation in the cross ratios. The current approach is equivalent. The analytic continuation is entirely captured by the choice of contour that defines $\E$ in the formulas above, implicit in the way we have ordered the operators, as we will discuss in detail in section \ref{s:ope}.

Conformal invariance was used several times in this derivation, but we expect the conclusions to apply also to non-conformal QFTs with an interacting fixed point in the UV. The approach to the lightcone is controlled by the UV fixed point, so if the fixed point is an interacting CFT, then the OPE formula \eqref{nulleope0} still applies. (This would not be true if the UV fixed point were free, since then an infinite tower of higher spin currents would contribute to the lightcone OPE at leading order.) One might worry that in the limit $u \gg 1$, some pairs of operators in the 4-point function are at large timelike separation, so perhaps there are significant infrared effects.  We do not have a complete argument that it is impossible, so leave this as an open question.  A similar OPE argument was used in \cite{Bousso:2014uxa} to derive Bousso's covariant entropy bound, and it was argued that such effects should be absent --- the same arguments apply here, so we consider this a mild assumption. See section \ref{ss:qfts} for further discussion.

\section{Average null energy in the lightcone OPE}\label{s:ope}

In this section, we will derive the universal contribution of the null energy operator $\E$, defined in \eqref{defE}, to $n$-point correlation functions in $(3+1)-$dimensions.  The general case ($d\neq 4$ and/or spin $> 2$) is in section \ref{GSO}.

\subsection{Lightcone OPE}

Consider a scalar primary $\psi$. In general, two nearby operators can be replaced by their operator product expansion,
\be
\psi(x_2)\psi(x_1) = \sum_i C_i^{\mu\nu\cdots}(x_1-x_2)O^i_{\mu\nu\cdots}(x_2) \ .
\ee
In the limit that $x_2-x_1$ becomes null, the OPE is organized as an expansion in twist, $\Delta_i - \ell_i$, where $\Delta$ is scaling dimension and $\ell$ is spin. For now we will assume that the stress tensor $T_{\mu\nu}$ is the unique operator of minimal twist. (This assumption is not necessary for the ANEC, as long as the theory is interacting and the stress tensor is the only spin-2 conserved current; see section \ref{GSO}.) Then the leading contributions to the OPE in the lightcone limit $v\to 0$ are
\be\label{lcope}
\psi(u,v)\psi(-u,-v) = \langle \psi(u,v)\psi(-u,-v) \rangle \left[1+ v u^3 \sum_{n=0}^{\infty} c_n (u\partial_u)^n T_{uu}(0) + \cdots \right] \ ,
\ee
with corrections suppressed by powers of $v$. We have inserted the operators symmetrically in the $uv$-plane with $\vec{x}=0$, and expanded about the midpoint. Other descendants of $T$ are subleading because they must come with powers of $v$ in order to contract indices.

The constants $c_n$ can be determined by plugging \eqref{lcope} into the 3-point function $\langle \psi \psi T_{\mu\nu}\rangle$ and comparing to the known answer, which can be found in \cite{Osborn:1993cr}.
However it is more elucidating to rewrite the lightcone OPE as an integral, rather than a sum. In fact \eqref{lcope} is exactly equivalent to
\be\label{enope}
\frac{\psi(u,v)\psi(-u,-v)}{\langle \psi(u,v)\psi(-u,-v) \rangle} = \left[1- \frac{15c_{\psi\psi T}}{ c_T} v u^2  \int_{-u}^u du' \left(1-\frac{{u'}^2}{u^2}\right)^2 T_{uu}(u',v=0)  + \cdots \right]\ .
\ee
This is derived by assuming an ansatz with an arbitrary kernel inside the integral, plugging into $\langle \psi\psi T_{\mu\nu}\rangle$, and designing the kernel to reproduce the known answer. Alternatively, we can expand the integrand as $T_{uu}(u') = T_{uu}(0) + u' \p_{u}T_{uu}(0) + \cdots$, do the integral, and check that \eqref{enope} reproduces \eqref{lcope} with the correct $c_n$'s. The OPE coefficient $c_{\psi\psi T}$ is fixed by the conformal Ward identity to $c_{\psi\psi T} = -\frac{2 \Delta_\psi}{3\pi^2}$.

In \eqref{enope}, the lightcone OPE is expressed as an integral of $T_{uu}$ over the null ray connecting the two $\psi$'s. It is an operator equation, meaning it can be used inside correlation functions, though we must be careful about convergence (or, equivalently, coincident point singularities).\footnote{See \cite{Czech:2016xec} for recent progress in writing general OPEs by integrating over causal diamonds. The OPE \eqref{lcope} can also be derived using shadow operators, as described in appendix B of that paper.  It is interesting to note the similarity to the formula for the vacuum modular Hamiltonian of an interval in 1+1 dimensions, $H \sim \int_{-L}^L (1-x^2/L^2)T_{tt}$,  and also to the recent derivation of the Bousso bound \cite{Bousso:2014uxa}, which relied on an integral expression for the null OPE of twist line operators.}

In the limit $u \to \infty$, the integration kernel is trivial, so the lightcone OPE produces the integrated null energy operator:
\be\label{nulleope}
\psi(u,v)\psi(-u,-v) \approx \langle \psi(u,v)\psi(-u,-v) \rangle(1+ \lambda_T vu^2 \E ) \ ,
\ee
where $\lambda_T =\frac{10 \Delta_\psi}{c_T \pi^2} $. This equation holds in the limit where we first take $v\to 0$, then $u\to \infty$, assuming that all other operator insertions are confined to some finite region.  Corrections are subleading in $1/u$ or $v$. 

\subsection{Contribution to correlators}
Now consider the correlator  
\be\label{pp1}
\langle \psi(x_1)\psi(x_2) O(x_3)\cdots O(x_n) \rangle \ .
\ee
($O$ may have spin; its indices are suppressed.) 

If all points are spacelike separated, then the $\psi\psi$ OPE is convergent. If, on the other hand, for instance $x_1 - x_3$ is spacelike but $x_2 - x_3$ is timelike, then the full $\psi\psi$ OPE may diverge. Still, we expect that any finite number of terms in the lightcone OPE produce a reliable asymptotic expansion in the limit $v_2 \to v_1$. This is argued in detail for 4-point functions in section 4.5 of \cite{causality1} by comparing to a different, convergent OPE channel. (More heuristically, it is reasonable to trust a divergent expansion in $v$ as $v \to 0$ so long as subsequent terms are highly suppressed, just as in ordinary perturbation theory.) For $n$-point functions, a similar argument holds. The conclusion is that \eqref{nulleope} can be used inside arbitrary correlation functions, as long as we take the limits with all other quantities held fixed. 

\begin{figure}
\begin{center}
\includegraphics{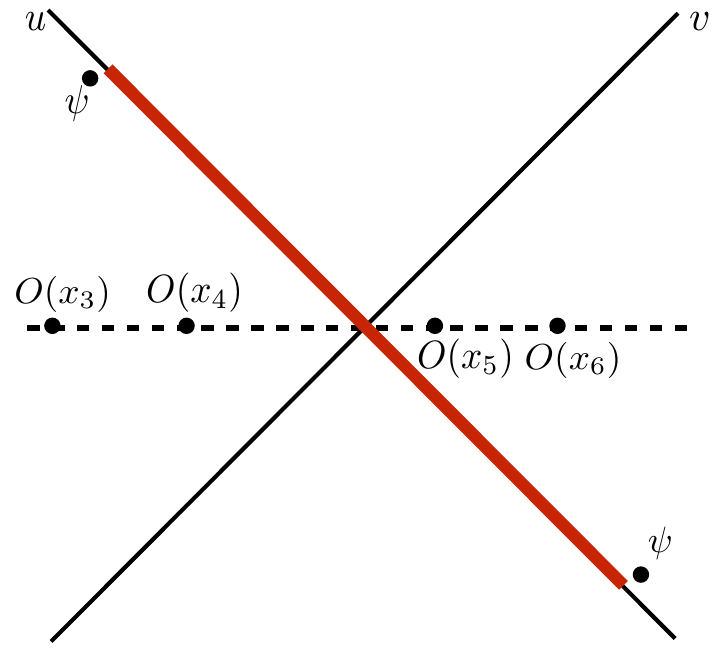}
\caption{\small Operator insertions in Minkowski spacetime. In the limit where the two $\psi$'s become null, but are widely separated in $u$, the leading non-identity term in the $\psi\psi$ OPE is $\int du T_{uu}$, integrated over the red null line.  \label{fig:oinsertions}}
\end{center}
\end{figure}

Operators are ordered by the standard prescription (reviewed in detail in \cite{causality1}):  To compute
\be\label{genoo}
\langle O_1(x_1)O_2(x_2)\cdots O_n(x_n)\rangle \ ,
\ee
shift $t_i \to t_i - i \epsilon_i$ with $\epsilon_1>\epsilon_2>\cdots>\epsilon_n$, and define the correlator by analytic continuation from the Euclidean. In the domain with a fixed imaginary time ordering, the function is analytic, and sending $\epsilon_i \to 0 $, it produces the Lorentzian correlator with operators ordered as written in \eqref{genoo}.  When we apply the lightcone OPE \eqref{nulleope}, this translates into a choice of contour for the $u$-integral. For concreteness, set $n=6$ and suppose the $O$'s are all at $t=\vec{x}=0$, with two $O$'s in each Rindler wedge, as in figure \ref{fig:oinsertions}.  (Generalizing to arbitrary Minkowski insertions  $x_{3,\dots,n} \in R^{1,d-1}$ with $x_i^2 > 0$ is straightforward.) Suppressing coordinates set to zero, first consider the ordering
\be\label{ppo1}
\langle \psi(u_1,v_1)\psi(-u_1,-v_1) O(y_3)O(y_4)O(y_5)O(y_6) \rangle 
\ee
with $v_1 < 0 < u_1$.
 In the limit $|v_1| \ll \frac{1}{|u_1|}\ll 1$, the OPE gives the leading terms
\be\label{ppo2}
 \langle \psi(u_1,v_1)\psi(-u_1,-v_1) \rangle \langle \left[1 + \lambda_T v_1 u_1^2 \int_{-\infty}^\infty du T_{uu}(u,v=0)\right]O(y_3)O(y_4)O(y_5) O(y_6)\rangle \ .
\ee
The integral has singularities at $-u = y_3,\dots,y_6$. The $i\epsilon$ prescription  says that to compute the correlator ordered as in \eqref{ppo1}, the contour in the complex $u$-plane goes below the poles:
\be\label{ucontour1}
\begin{gathered}\includegraphics{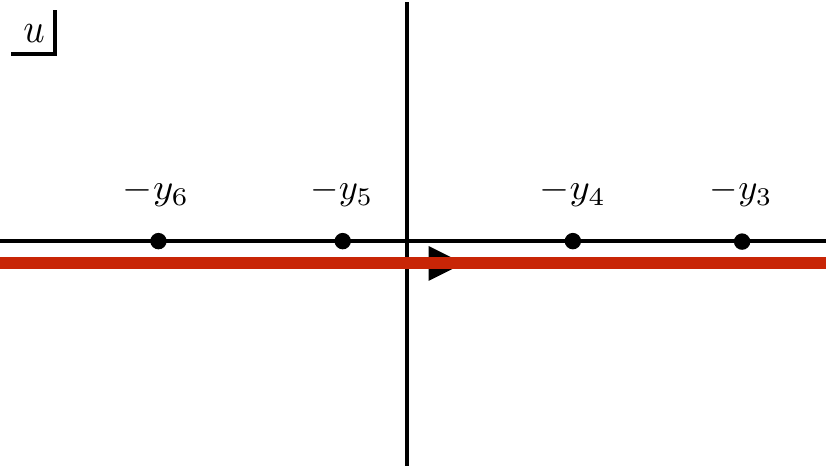}\end{gathered} 
\ee
As $|u|\to\infty$, applying the OPE to all the $O$'s implies that the integrand falls off the same as (or faster than) $\langle T_{uu}(u)T_{\alpha\beta}(x)\rangle \sim u^{-6}$, so we can deform the contour and the integral vanishes. (This does not mean that the stress tensor contribution to the lightcone OPE vanishes, only that it has no terms $\sim v_1 u_1^2$. The first non-zero contribution is actually $\sim v_1/u_1^3$, using \eqref{enope}.)
Other orderings are obtained by deforming the contour across poles. For example, the time-ordered correlator
\be
\langle \psi(u_1,v_1) O(y_3)O(y_4)O(y_5)O(y_6)\psi(-u_1,-v_1) \rangle
\ee
is again computed by \eqref{ppo2}, but now integrating on the following contour:
\be\label{ucontour2}
\begin{gathered}\includegraphics{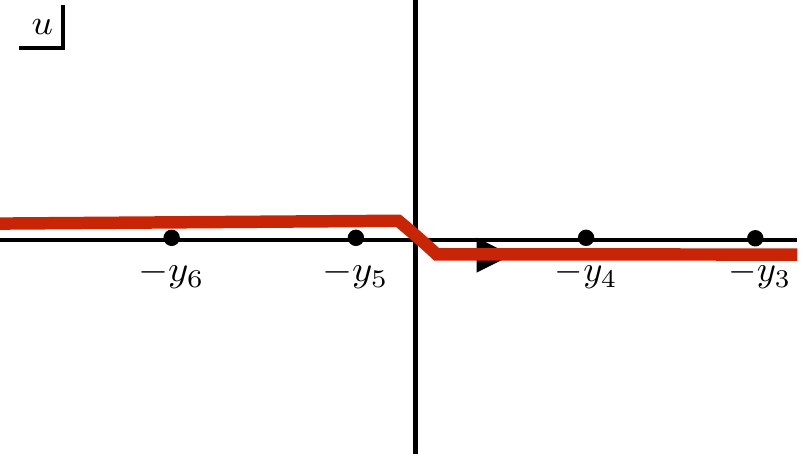}\end{gathered}
\ee
The integral is equal to the sum of residues at $u=-y_5, -y_6$ or at $u = -y_4, -y_3$, so this ordering does have terms $\sim v_1 u_1^2$.

\subsection{Scalar example}
In the language of \cite{mss,causality1}, the trivial contour \eqref{ucontour1} is the 1st sheet (or Euclidean) correlator, while the non-trivial contour \eqref{ucontour2} produces the correlator on the 2nd sheet.

As a simple application, let us reproduce the well known hypergeometric formula for the four-point conformal block in the lightcone limit.  Consider the four-point function of identical scalars, 
\be
G_{scalar}(z,\bz) = \langle \phi(0) \phi(z,\bz) \phi(1) \phi(\infty)\rangle \ .
\ee
With these kinematics, the cross ratios are simply the lightcone coordinates of the second insertion. For $z,\bz \in (0,1)$, all points are spacelike.  Plugging in the lightcone OPE \eqref{enope} gives an integral that is easily recognized as a hypergeometric function, and we reproduce the well known formula for the stress tensor lightcone block (see \cite{Komargodski:2012ek,Fitzpatrick:2012yx}):
\be
 \bz z^3 \,_2F_1(3,3,6,z) = -\frac{30 \bz}{z^2}\left[-3(-2+z)z + (6-6z+z^2)\log(1-z)\right] \ .
\ee
This is regular as $z \to 0$.  But after going to the second sheet, \ie sending $\log(1-z) \to \log(1-z) - 2\pi i$, the behavior near the origin is $\sim i \bz/z^2$.  This growing term, with the correct coefficient, is what is captured by the approximation where we replace the full lightcone OPE by just the null energy operator, as in \eqref{nulleope}. So what we have shown is that these growing terms, responsible for all the results in \cite{causality1,causality2,Hofman:2016awc}, are precisely the contributions of the null energy operator $\E$.

\section{Sum rule for average null energy}\label{s:properties}

Now we will fill in the details of the ANEC derivation in section \ref{s:anec}. Most of this discussion is a review of \cite{mss} and \cite{causality1}, some of it from a different perspective. First, we will collect some facts about position-space correlation functions, which hold in any unitary, Lorentz-invariant QFT. Then we put them together to derive the sum rule and positivity condition.

\subsection{Analyticity in position space}

The point of view taken throughout the paper is that a QFT can be defined by its Euclidean $n$-point correlation functions \cite{haag}.  These are single-valued, permutation invariant functions of $x_{1,\dots,n} \in R^d$, \ie there are no branch cuts in Euclidean signature. This ensures that in the Lorentzian theory, non-coincident local fields at $t=0$ commute with each other.

The Euclidean correlators can be analytically continued to complex $x_i$.  However, there are branch points when one operator hits the lightcone of another. (See \cite{haag} for details or section 3 of \cite{causality1} for a review.) When we encounter one of these branch points, we must choose whether to go around it by deforming $t \to t + i \epsilon$ or $t \to t-i\epsilon$, and this selects whether the two operators are time-ordered or anti-time-ordered. Thus the $i\epsilon$ prescription to compute a Minkowski correlator ordered as
\be
\langle \phi_1(x_1) \phi_2(x_2) \phi_3(x_3) \cdots\rangle
\ee
is to gives each $t_i$ a small imaginary part, with 
\be\label{iepo}
\Im t_1 < \Im t_2 < \Im t_3 <  \cdots \ .
\ee
The resulting function is analytic as long as we maintain \eqref{iepo}, so once we've specified the $i\epsilon$ prescription, the analytic continuation from Euclidean is unambiguous. 

In fact, the domain of analyticity of the $n$-point correlator $G(x_i)$, viewed as a function on (a subdomain of a cover of) $n$ copies of complexified Minkowski space, is larger than indicated by \eqref{iepo}. It is also analytic on the domain defined by the covariant version of \eqref{iepo}:
\be\label{copo}
\Im x_1  \vartriangleleft \Im x_2 \vartriangleleft \Im x_3 \vartriangleleft \cdots ,
\ee
where $x \vartriangleleft y$ means `$x$ is in the past lightcone of $y$'.\footnote{We define the imaginary part of a complexified point by the convention that a point in Minkowski space $R^{1,d-1}$ has $x_i = \mbox{Re\ } x_i$. Thus the real and imaginary parts each live in a copy of Minkowski space, not Euclidean space.
} (Actually, it is analytic on an even larger domain, but \eqref{copo} is all we need.\footnote{The full domain is described as follows \cite{haag}. First act on the domain \eqref{copo} by all possible complex Lorentz transformations; this defines the \textit{extended tube}. Then, permute the $n$ points and a repeat this procedure, to define the \textit{permuted extended tube}. In the theory of multiple complex variables, the domain of analyticity cannot be an arbitrary shape --- once we know a function is analytic on some domain, it must actually be analytic on a (generally larger) domain called the \textit{envelope of holomorphy}. The domain of analyticity of the correlator $G(x_i)$ is the envelope of holomorphy of the union of permuted extended tubes.})

\subsection{Rindler positivity}\label{RP}
Correlation functions in Minkowski space restricted to the left and right Rindler wedges obey a positivity property analogous to reflection positivity in Euclidean signature. This is derived in \cite{Casini:2010bf}\footnote{What we call Rindler positivity is not, however, quite the same as `wedge reflection positivity' referred to in the title of the paper \cite{Casini:2010bf}. The difference is discussed in section 3 of \cite{Casini:2010bf}.} and also in \cite{mss}.

Define the Rindler reflection
\be
\overline{x} = \overline{(t,y,\vec{x})} = (-t^*, -y^*, \vec{x}) \ .
\ee
The transverse coordinate $\vec{x}$ is taken to be real. For real $(t,y)$, this reflects a point in the right Rindler wedge to the left Rindler wedge. Acting on a spinning operator $O$,
\be
\overline{O_{\mu\nu\cdots}(t,y,\vec{x})} = (-1)^P O^\dagger_{\mu\nu\cdots}(-t^*, -y^*, \vec{x})
\ee
where $P$ is the number of $t$-indices plus $y$-indices.  (The Hermitian conjugate on the right-hand side acts only on the operator, not the coordinates.) This operation is CPT together with a rotation by $\pi$ around the $y$-axis: $\overline{O} = J O J $ with $J = U(R(y, \pi))CPT$ \cite{Casini:2010bf}.

We will insert points in a complexified version of the left and right Rindler wedges. The complexified right wedge is defined as
\be
R_C = \{ (u,v,\vec{x}) : uv<0, \ \mbox{arg\ }v \in (-\frac{\pi}{2}, \frac{\pi}{2}) ,\ \vec{x} \in R^{d-2} \} \ ,
\ee
and the complexified left wedge is
\be
L_C = \overline{R_C} = \{ (u,v,\vec{x}) : uv<0, \ \mbox{arg\ }v \in (\frac{\pi}{2}, \frac{3\pi}{2}) ,\ \vec{x} \in R^{d-2} \} \ .
\ee
The positivity condition for $2n$-point correlators is
\be\label{corpos}
\langle \overline{O_1(x_1)}\,  \overline{O_2(x_2)} \, \cdots \overline{O_n(x_n)} O_1(x_1)O_2(x_2)\cdots O_n(x_n)\rangle > 0 \ ,
\ee
for $x_{1,\dots,n} \in R_C$ and
\be\label{imorder}
\Im t_1 \leq \Im t_2 \leq \cdots \leq \Im t_n \ .
\ee
Note that for product operators, the order does not reverse under reflection:
\be
 \overline{O_1 O_2}=\overline{O_1} \ \overline{O_2}  \ .
\ee
For real insertions, $\Im t_i = 0$, the operators in \eqref{corpos} are ordered as written, which we will refer to as `positive ordering'. They are not time ordered. For complex insertions, the correlator is defined by analytic continuation from Euclidean within the domain \eqref{iepo}.  The reflected operators have
\be
\Im -t_1^* \leq \Im -t_2^* \leq \cdots \leq \Im -t_n^* \ ,
\ee
which explains why they must be ordered as in \eqref{corpos}.

Positivity applies also to smeared operators, and products of smeared operators, with support in a single complexified Rindler wedge. That is,
\be
\langle \overline{\Theta} \Theta \rangle > 0 \ ,
\ee
for
\be
\Theta = \int f^{(1)}(x_1)O_1(x_1) + \int \int f^{(2)}(x_1,x_2)O_1(x_1)O_2(x_2) + \cdots \ ,
\ee
where the smearing functions $f$ have support in some localized region of $R_C$, and the operator ordering in $\overline{\Theta}$ is the same as in $\Theta$.

\begin{figure}
\begin{center}
\includegraphics{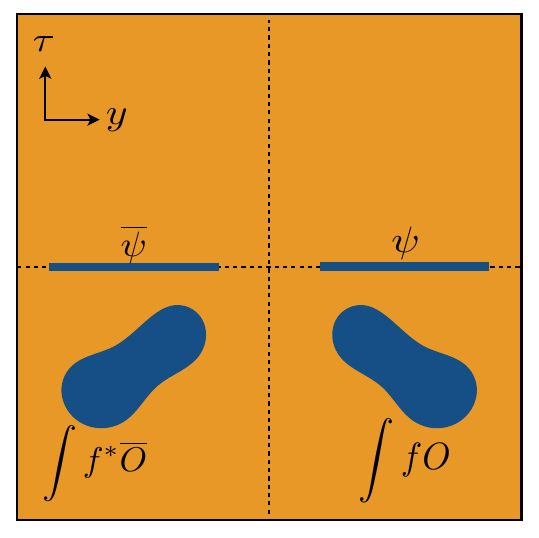}
\end{center}
\caption{\small Operator insertions on the $\tau y$-plane defining the smeared 4-point function $\langle \bTheta_0 \Theta_0 \rangle$.\label{fig:reflection}}
\end{figure}

These positivity conditions hold in any unitary, Lorentz-invariant QFT \cite{Casini:2010bf,mss}. We will not repeat the derivation, but an intuitive way to understand this is as follows. To be concrete, consider a case of particular interest for the ANEC and the Hofman-Maldacena constraints that will be used below:
\be
\Theta_0 = \psi(t_0,y_0) \int_{0}^\infty d\tau \int_0^{\infty}dy \int d^{d-2}\vec{x} f(\tau, y, \vec{x}) O(t=-i\tau, y, \vec{x})
\ee
with $t_0, y_0 >0$, and assume $f$ is non-zero in some finite region.  $\psi$ and $O$ may be timelike separated, in the sense that Re $(x_0 - x) \in R^{1,d-1}$ lies in the forward lightcone for points of the integral. We want to understand why $\langle \bTheta_0 \Theta_0\rangle > 0$. First, we can evolve $\psi(t_0,y_0)$ back to the $t=0$ slice; it becomes non-local, but with support only in the right Rindler wedge.  The same can be done in $\bTheta_0$, evolving $\psi(-t_0,-y_0)$ forward to the $t=0$ slice. Then in the $\tau y$-plane, the 4-point function $\langle \bTheta_0 \Theta_0\rangle$ is smeared over the regions shown in figure \ref{fig:reflection}.  These insertions are symmetric under $y \to -y$ together with complex conjugation; therefore, reflection positivity of the Euclidean theory guarantees that this correlator is positive. Keeping track of Lorentz indices on $O$ leads to the same conclusion. For a more precise derivation we refer to \cite{Casini:2010bf} for real insertions, and \cite{mss} for insertions in the complexified wedge $R_C$. 

\begin{figure}
\begin{center}
\includegraphics{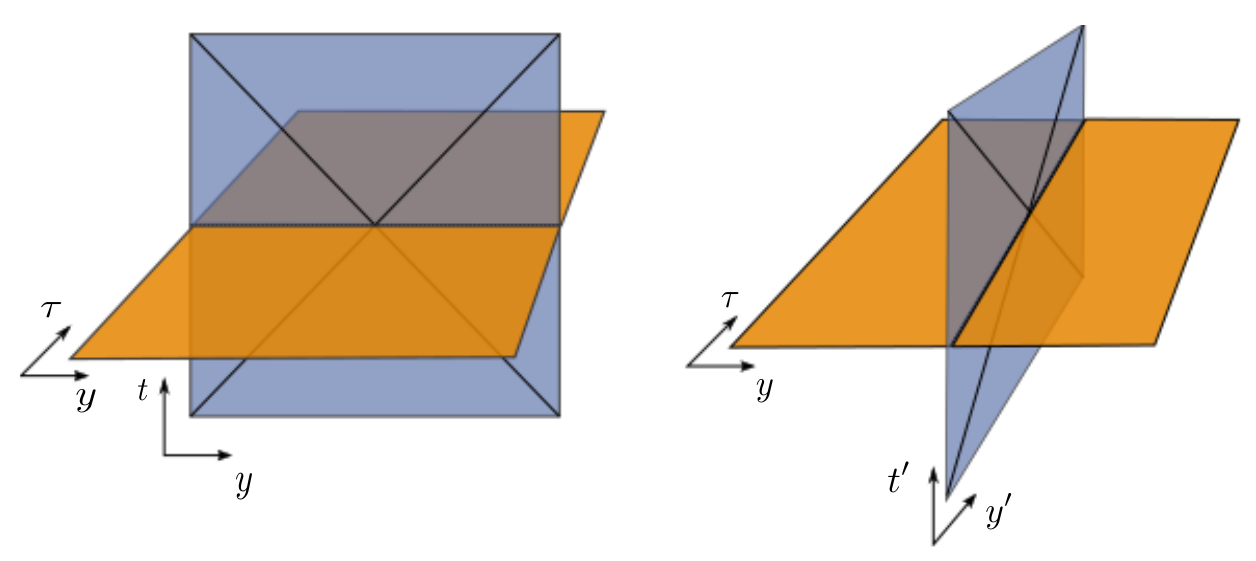}
\end{center}
\caption{\small Two different ways to interpret the same Euclidean theory. The Euclidean $R^d$ (horizontal orange plane, parameterized by $(\tau,y,\vec{x})$) is the same in both pictures, but the definition of states and corresponding notion of Minkowski spacetime (vertical blue planes) is different in the two cases. On the left, the continuation to Lorentzian is $\tau \to i t$, states of the theory are defined on the plane $\tau = 0$, and $y$ is a space direction. On the right, the continuation to Lorentzian is $y \to i t'$, states are defined at $y = 0$, and $\tau=y'$ is a space direction. The two theories are identical, since they are determined by the same set of Euclidean correlators, but the map of observables and matrix elements from one description to the other is nontrivial.\label{fig:minkowski1}}
\end{figure}

To recap, although $\langle \bTheta \Theta\rangle$ does not look like a norm in the theory quantized on the $\tau=0$ plane --- since $\Theta |0\rangle$ is not normalizable, and $\bTheta \neq \Theta^\dagger$ --- it is a norm in the theory quantized on the $y=0$ plane.  These two different quantizations correspond to two different ways of analytically continuing a Euclidean theory to Minkowski space as shown in figure \ref{fig:minkowski1}. 

In conformal field theory, the positivity properties discussed here follow from the fact that the conformal block expansion has positive coefficients (in the appropriate channel), as described in \cite{causality1,causality2}. We have chosen a different but conformally related kinematics in the present paper because (i) it makes the positivity conditions more manifest, (ii) positivity in the new kinematics does not require conformal invariance, and (iii) it allows us to smear operators while easily maintaining positivity properties needed to derive the  constraints.

\subsection{Bound on the real part}
With operators inserted symmetrically across the Rindler horizon, the positive-ordered correlator is real and positive. The time-ordered correlator is generally complex, but it inherits from \eqref{corpos} bounds on its magnitude and real part. This was derived in \cite{mss} by interpreting the correlator as a Rindler trace and applying Cauchy-Schwarz inequalities. The positivity condition for CFT shockwaves derived in \cite{causality1,causality2} using the decomposition into conformal partial waves can also be restated in this way.

Here we repeat the Cauchy-Schwarz derivation, but in Minkowski language. Let $A$, $B$ be operators (possibly nonlocal) with support in the right wedge $R_C$. The positive-ordered correlator defines a positive inner product $(A,B) \equiv \langle \overline{A} B \rangle$. Therefore the Cauchy-Schwarz inequality applies,
\be
|\langle \overline{A} B \rangle|^2 \leq \langle \overline{A} A\rangle \langle \overline{B} B\rangle \ .
\ee
In the derivation of the ANEC in section \ref{s:anec} we considered
\be
G_{anec} = \langle  \overline{O}\psi(u,v) \psi(-u,-v) O \rangle \ .
\ee
(There $O$ was local, but for the present purposes it can also be smeared.) Applying Cauchy-Schwarz with $A=O\psi(-u,-v)$, $B = \psi(-u,-v) O$,
\renewcommand{\bO}{\overline{\mathcal{O}}}
\newcommand{\bpsi}{\overline{\psi}}
\be\label{reabs}
\Re G_{anec} \leq |G_{anec}| \leq \left( \langle \bO \bpsi O \psi\rangle \langle \bpsi \bO \psi O\rangle \right)^{1/2}
\ee
where $\psi \equiv \psi(-u,-v), \bpsi \equiv \psi(u,v)$. Note that both of the correlators on the right-hand side are positive-ordered.

\subsection{Factorization}
In the limit $u \to \infty$ (with everything else held fixed or $v \sim 1/u$), positive-ordered correlators factorize into products of two-point functions. In a CFT, this follows from the OPE. In this limit, we can replace $O$ by a local operator, and the conformal cross ratios of the 4-point function $ \langle \bO \bpsi O \psi\rangle  $ are $z,\bz \sim 0$. Positive ordering means that we do not cross any branch cuts to reach this regime \cite{mss,causality1}, so the correlator can be computed by the usual Euclidean OPE and is dominated by the identity term. Thus
\be
 \langle \bO \bpsi O \psi\rangle \sim \langle \bpsi \bO \psi O\rangle \sim \langle \bO O\rangle \langle \bpsi \psi\rangle
\ee
and \eqref{reabs} becomes
\be\label{rebanec}
\Re G_{anec} \leq |G_{anec}| \leq \langle \bO O\rangle \langle \psi(u,v)\psi(-u,-v)\rangle + \varepsilon \ .
\ee
The correction term $\varepsilon$ on the right is necessary because the positive-ordered correlator does not exactly factorize. It has corrections from subleading operator exchange. But $\varepsilon$ is suppressed by positive powers of $v$ and $1/u$, so it can be neglected everywhere in the derivation.

This argument relied on conformal invariance, but factorization is expected to hold in a general interacting theory (possibly excluding integrable theories).  This is motivated on physical grounds in \cite{mss} by relating it to thermalization of finite temperature Rindler correlators.

\subsection{Sum rule}\label{ss:sumrule}
Now we use the method of \cite{causality1} to derive a manifestly positive sum rule for averaged null energy, $\langle O | \E | O\rangle$. Let
\be
v = -\eta \sigma , \quad u = 1/\sigma \ ,
\ee
with $0 < \eta \ll 1$ and consider the function $G$ defined in \eqref{maing} as a function of complex $\sigma$, with all other coordinates fixed. The function $G(\sigma)$ obeys two important properties:
\begin{enumerate}[label=(\roman*)]
\item \label{analytic}  $G(\sigma)$ is analytic on the lower-half $\sigma$ plane in a region around $\sigma \sim 0$. This follows from \eqref{copo} with complexified points labeled as $x_1 = (-u,-v)$, $x_2 = (y=\delta)$, $x_3 = (y=-\delta)$, $x_4 = (u,v)$.
\item \label{realbound} For real $\sigma$ with $|\sigma| < 1$, 
\be\label{regbound}
\mbox{Re}\ G \leq 1 + \varepsilon \ ,
\ee
where the correction $\varepsilon$ is suppressed by positive powers of both $\eta$ and $\sigma$. This follows directly from \eqref{rebanec}.
\end{enumerate}
Equipped with these facts, the sum rule is derived by integrating $G(\sigma)$ over the boundary of a half-disk, just below $\sigma = 0$:
\be\label{contour}
\includegraphics{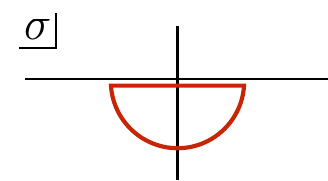} 
\ee
The radius $R$ of the semicircle does not matter, as long as it is in the regime \eqref{uvlim}, \textit{i.e.,} $\eta \ll R \ll 1$. The integral over a closed contour vanishes,
\be
\mbox{Re} \oint d\sigma (1-G(\sigma)) =0 \ .
\ee
We split this into the contribution from the semicircle and from the real line, then use the OPE \eqref{gcor} to evaluate the correlator on the semicircle and do the integral. The result is
\be
i \langle  \overline{O(y=\delta)} \, \E \, O(y=\delta)= \frac{N_\delta}{\pi \eta \lambda_T} \int_{-R}^R d\sigma \mbox{Re}\ (1-G(\sigma)) \ .
\ee
The right-hand side is positive by property \ref{realbound} above. Using \eqref{orotate}, the sum rule can be written as a manifestly positive integral for the expectation value of average null energy:
\be\label{mainsum}
\langle O| \E |O\rangle  = \frac{1}{\pi \eta \lambda_T} \int_{-R}^R \Re(1 - G(u=\frac{1}{\sigma}, v= - \eta \sigma))
\ee
where $|O\rangle \equiv \frac{1}{\sqrt{N_\delta}} O(t=-i\delta)|0\rangle $ (and if $O$ is not a scalar, the operator is rotated by $\pi/2$ in the Euclidean $\tau y$-plane).

Note that two distinct positivity conditions came into play.  First, there was Rindler positivity, property \ref{realbound}. Applied to the correlator $G$ in the lightcone regime, this would imply $\Re \langle \overline{ O(y=\delta)} \E O(y=\delta) \rangle \geq 0$. 
However, this 3-point function is purely imaginary, so this constraint is trivial. Rindler positivity is nontrivial only near the origin of the $\sigma$-plane (the Regge-like limit), where the OPE is not dominated by the low-dimension (or low-twist) operators.  Second, there is the positivity condition on the imaginary part of the OPE correction, coming from the sum rule.  It is this second, less direct consequence of reflection positivity that leads to the ANEC. This is similar to the use of dispersion relations for scattering amplitudes in momentum space -- the optical theorem gives a positivity condition on the forward amplitude, and sum rules relate this positive quantity to the amplitude in other regimes (see for example \cite{Adams:2006sv}).

\subsection{Non-conformal QFTs}\label{ss:qfts}

As mentioned in section \ref{s:anec}, we expect the main conclusions and the sum rule to hold even in non-conformal QFTs as long as there is an interacting UV fixed point. This is essentially because it is a UV argument. The applicability of the lightcone OPE was discussed at the end of section \ref{s:anec}. Conformal symmetry was used again when we invoked the state-operator correspondence to claim that any normalizable state can be created by inserting a local operator at $t = -i\delta$. This made the derivation simpler, since we could restrict to local operator insertions $O(y=\delta, t=0, \vec{x}=0)$. But in fact this restriction was not necessary. The ingredients that go into the sum rule --- positivity, etc --- still apply if $O$ is a non-local operator defined by smearing local operators (and their products) over a complexified version of the Rindler wedge.  In any QFT a dense set of states can be created in this way \cite{haag}.

\section{Hofman-Maldacena bounds}\label{s:hm}
The conformal collider bounds of Hofman and Maldacena \cite{Hofman:2008ar} are constraints on CFT 3-point functions that come from imposing
\be\label{hmpos}
\lim_{v \to \infty} v^2 \langle \e^* \sdot O(P)^\dagger \int_{-\infty}^\infty du T_{uu}(u,v) \e \sdot O(P)\rangle \geq 0 \ .
\ee
Here $O(P)$ is a wave packet with timelike momentum $P = \omega \hat{t}$, created by inserting a spinning operator near the origin: 
\be
\e \sdot O(P) = \int dt dy d^{d-2}\vec{x}\, e^{-(t^2 + y^2 + \vec{x}^2)/D^2} e^{-i\omega t}\e^{\mu\nu\cdots}O_{\mu\nu\cdots}(t,y,\vec{x}) \ , \qquad \omega D \gg 1 \ .
\ee 
The positivity condition \eqref{hmpos} was an assumption in \cite{Hofman:2008ar}, motivated by the fact that this computes the energy measured in a far-away calorimeter if we prepare a CFT in the state $\e \sdot O(P) | 0\rangle$. It leads to constraints on the 3-coupling constants that appear in $\langle OTO\rangle$.

Since we have shown that $\E$ is a positive operator, the inequality \eqref{hmpos} follows from the above analysis. But it is instructive to see how constraints in this particular state are related to our discussion of Minkowski scattering and the ANEC sum rule.  That is the goal of this section.  In particular, we will show exactly how to smear the probe operators in the previous analyses \cite{causality1,causality2,Hofman:2016awc} to produce the Hofman-Maldacena inequalities. This avoids the step of decomposing the correlator into the crossed channel, used in \cite{Hofman:2016awc} in order to improve upon the bounds derived in \cite{causality2}.

\subsection{Conformal collider redux}
First we will restate the Hofman-Maldacena condition in a way that makes all of the integrals trivial. We perform the Fourier transform over $t$ first. In the regime $\omega D \gg 1$ it is dominated by a saddlepoint at $t = -\tfrac{i}{2}\omega D^2$. Therefore, instead of viewing this as a wavepacket with frequency $\omega$, we can replace it by an operator inserted at a fixed, imaginary value of $t$:
\be\label{gauso}
\e \sdot O(P) \approx \int dy d^{d-2}\vec{x} \, e^{-(y^2 + \vec{x}^2)/D^2}  \e^{\mu\nu\cdots}O_{\mu\nu\cdots}(t=-i\delta,y,\vec{x}) 
\ee
with $\delta >0$. Also, in this limit, we only need to integrate over the position of one of the $O$ insertions, since the other integral gives an overall factor.  The remaining gaussian can be dropped, and the final $d-1$ integrals are done by residues.\footnote{Dropping the gaussian can lead to unphysical divergences in the remaining integrals, depending on the dimensions of the operators. These are dealt with by dimensional regularization in the transverse directions, $d-2 \to d- 2 + \epsilon$ with $\epsilon \to 0$ at the end \cite{Chowdhury:2012km}.} 

In \eqref{hmpos}, the state is created near the origin of Minkowski space, and the average null energy is evaluated near future null infinity. For comparison to the rest of the paper, it is more convenient to shift coordinates so that the null energy is integrated over a ray at $v=0$, and $O$ is inserted near spatial infinity. That is, \eqref{hmpos} is equivalent to 
\be\label{hmshift}
\lim_{\lambda \to \infty} \lambda^2 \langle \varepsilon \sdot O_\lambda| \, \E \, | \varepsilon \sdot O_\lambda \rangle \geq 0
\ee
where $\E$ is integrated over $v=0$ as in \eqref{defE}, and
\be\label{lamo}
|\varepsilon \sdot O_\lambda\rangle \equiv \int d\tilde{y} d^{d-2}\vec{x}\, \varepsilon\sdot O(t=-i\delta, y = \lambda \tilde{y}, \vec{x}) |0\rangle \ .
\ee
In \eqref{hmshift}, the wavepacket is implemented by the order of limits: first we do the $u$-integral (by residues) to compute $\E$, then take $\lambda \to \infty$, then perform the integrals over $\vec{x},\tilde{y}$. 

To recap, the Hofman-Maldacena constraints are restated as:
\be\label{finalhm}
 \int d\tilde{y} d^{d-2}\vec{x}\,  \lim_{\lambda \to \infty} \lambda^2 \langle \varepsilon^* \sdot O(t=i\delta, y=\lambda,\vec{0}) \, \E  \,
 \varepsilon\sdot O(t=-i\delta, y = \lambda \tilde{y}, \vec{x})\rangle \geq 0 \ .
\ee
This expression is a convenient way to compute the explicit constraints in terms of the 3-point coupling constants.

\subsection{Relation to scattering with smeared insertions}
The state \eqref{lamo} is most naturally created by smearing an operator near spatial infinity, but like any localized state in a CFT, it can also be created by inserting a single, non-primary operator at $t=-i\delta, y=\vec{x} = 0$.\footnote{Everywhere in this section, the limit $\lambda \to \infty$ should be interpreted as large but finite $\lambda$, with $|v| \ll \frac{1}{|u|} \ll \frac{1}{\lambda} \ll 1$ and $\delta \sim 1$.   As we move the wavepacket further, the constraints obtained this way approach arbitrarily close to the Hofman-Maldacena type constraints.  (And, in the version of the argument with smeared insertions, we put a hard cutoff on the wavepacket at some distance where it is exponentially suppressed, in order to avoid coincident point singularities.)}  It is straightforward to find the operator explicitly by a series expansion of the wavepacket integral, $\int dy d^{d-2}\vec{x}\,  O(t=-i\delta,y,\vec{x})e^{-((y-\lambda)^2+\vec{x}^2)/D^2}$. 
Applying section \ref{s:anec} to this particular operator is one way to derive the Hofman-Maldacena inequalities directly from the causality of the 4-point function.  However, we would need to be more careful about the order of limits in the series defining the operator and various other steps of the calculation, especially since we are expanding a wavepacket about a point very far from its center. 

This problem is avoided if we instead apply the causality argument directly to a correlator with smeared operator insertions, corresponding to wavepackets offset far into imaginary time. As explained around figure \ref{fig:minkowski1}, the interpretation of the ANEC as an expectation value and the interpretation of the ANEC as it appears in the lightcone OPE differ by a $\pi/2$ rotation in the Euclidean $\tau y$-plane.  Therefore, after the rotation, in order to make contact with section \ref{s:anec} we are led to study the correlator
\be\label{ghm}
G_{HM} = \frac{\langle \overline{O_{HM}} \psi(u,v) \psi(-u,-v) O_{HM}\rangle}{ \langle \overline{O_{HM}} O_{HM}\rangle \langle \psi(u,v) \psi(-u,-v)  \rangle}
\ee
where
\be
O_{HM} = \lim_{\lambda \to \infty} \int d\tau d^{d-2}\vec{x}\ \varepsilon \sdot O(t=i\tau,y=\delta,\vec{x}) e^{-((\tau-\lambda)^2 + \vec{x}^2)/D^2}
\ee
The insertions are now symmetric under Rindler reflection, and the wavepacket is centered at large imaginary time. We could of course map the centers to real points in Minkowski, but the smearing procedure in this frame would be more complicated.

As in the discussion of the Hofman-Maldacena calculation above, the wavepacket can be implemented by instead taking
\bea
O_{HM} &=& \int d\tau d^{d-2}\vec{x}\  \varepsilon\sdot O(t=-i \lambda+i \tau, y=\delta, \vec{x})\ , \\
\overline{O_{HM}} &=& \int d\tau' d^{d-2}\vec{x}' \ \overline{\varepsilon}\sdot O^\dagger(t=-i \lambda+i \tau', y=-\delta, \vec{x}')\ ,
\eea 
where $\overline{\varepsilon}^{\mu\nu}=(-1)^P(\varepsilon^{\mu\nu})^*$ and $P$ is the number of $t$-indices plus $y$-indices; the integral in $\E$ is done first, then the limit $\lambda \to \infty$, then the remaining integrals.

The leading correction to \eqref{ghm} in the lightcone limit comes, as usual, from the integrated null energy:
\be\label{dg}
\delta G \sim \langle \overline{O_{HM}}\,  \E \, O_{HM} \rangle \ .
\ee
The operator ordering here and in \eqref{ghm} is subtle:  we do \textit{not} follow the usual prescription of analytic continuation \eqref{iepo} which orders operators by imaginary time. If we did, then the $u$-contour (in the integral defining $\E$) would not circle any poles, and the constrained term would vanish as in \eqref{ucontour1}.  Instead we define the correlator in \eqref{ghm} with the $u$-contour as follows:
\be\label{hmc}
\begin{gathered}\includegraphics[width=200px]{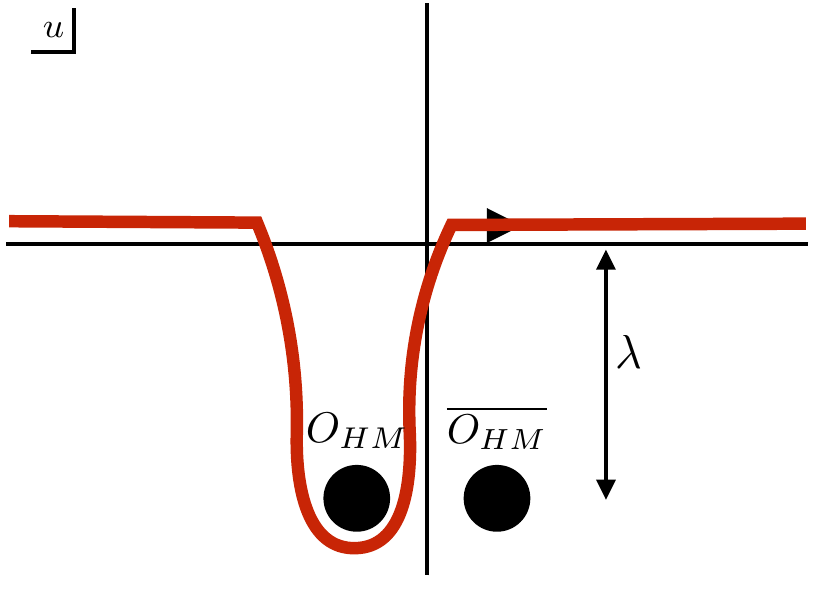}\end{gathered}
\ee
The black circles indicate the wavepacket insertions.  This picture can be interpreted two ways: First, it shows the path of analytic continuation that defines $G_{HM}$, ie the route taken by the $\psi$'s as we push $\psi(u,v)$ forward in time starting from the Euclidean correlator.  Second, it is the contour of integration used when $\E$ appears in the OPE. \footnote{An equivalent prescription is to define the correlator by first expanding each wavepacket in a series expansion around $t=0$, then compute the usual time-ordered correlator. Yet another equivalent prescription is to first compute the time-ordered correlator at $\lambda \sim 0$ as we did in previous sections, then analytically continue to finite $\lambda$.} The correlator defined in this way has the same properties as $G_{anec}$ in section \ref{s:anec}, so the sum rule \eqref{mainsum} still applies; thus the correction \eqref{dg} has a positive imaginary part, and this inequality is identical to the conformal collider constraints.

\subsection{Relation to the shockwave kinematics}
In earlier work \cite{causality1,causality2,Hofman:2016awc} a different kinematics was used for the four-point function in order to derive causality constraints. There, the kinematics corresponded to expectation values of $OO$ in a shockwave state created by $\psi$ inserted near the origin.  Although the conformal cross ratios are the same in the two setups, the advantage of the scattering kinematics used here is that positivity conditions are now manifest, using Rindler reflections as in \cite{mss}. This makes it easy to generalize the argument to non-primary, smeared insertions, which does not appear to be straightforward using the kinematics of \cite{causality1,causality2,Hofman:2016awc}.

\section{New constraints on higher spin operators}\label{GSO}

So far we have discussed constraints on the integrated stress tensor.  As in many other contexts (for example \cite{Adams:2006sv,Komargodski:2012ek,causality1,Komargodski:2016gci,causality2,Hofman:2016awc}, the positive sum rules for spin-2 exchange readily generalize to the exchange of higher spin operators. Let $X$ be the lowest-dimension operator of spin $s$, where $s > 2$ is even. This operator is the dominant spin-$s$ exchange in the lightcone limit \cite{Komargodski:2012ek}. We will argue that
\be\label{defes}
\E_s = \int_{-\infty}^\infty du X_{uuu\cdots u}(u,v=0,\vec{x}=0)
\ee
is positive in any state. 
The resulting constraints agree in many cases with other methods, but are generally stronger for non-conserved operators. It would be  interesting to check them in known conformal field theories, for example by numerical bootstrap or other methods. Although our OPE method does not apply to free theories, it is shown by direct calculation in appendix \ref{s:free} that \eqref{defes} holds for free scalars.

The formulas in this section also hold for $s=2$, so this generalizes the discussion in the rest of the paper to the ANEC in any spacetime dimension $d>2$.

\subsection{$\E_s$ in the lightcone OPE}

First let us derive the contribution of the operator $X$ and its descendants to the operator product expansion of two scalars in the lightcone limit. Repeating the steps in section \ref{s:ope}, we find the lightcone OPE can be written as the integral
\begin{align}\label{ope1}
\psi(u,v)\psi(-u,-v)|_{X} &= \langle \psi(u,v)\psi(-u,-v) \rangle  \frac{2^{\Delta_X} c_{\psi\psi X}  \Gamma \left(\frac{\Delta_X+s+1}{2}\right)}{ \sqrt{\pi}c_X  \Gamma \left(\frac{\Delta_X+s}{2}\right)}\frac{ (-v)^{\tau_X/2 }}{ {u}^{\frac{\Delta_X+s}{2}-1} }\nn\\
& \quad \quad \times \int_{-u}^{u}du' \left(u^2-u'^2\right)^{\frac{\Delta_X+s}{2}-1} X_{uu\cdots u}(u', 0)
\end{align}
where $c_{\psi\psi X}$ is the OPE coefficient, $c_X>0$ is the coefficient of the $XX$ two-point function, and $\tau_X = \Delta_X - s$ is the twist. (Conventions for the constants follow \cite{Costa:2011mg}). The notation $(\cdots)|_X$ means the contribution of the operator $X$, and the expression in \eqref{ope1} is the full contribution of the operators $(\p_u)^n X_{uu\cdots u}(0)$ for $n \geq 0$. Other descendants are subleading in the lightcone limit.

The OPE in the regime relevant to the ANEC is a divergent asymptotic series, organized by twist $\tau$ schematically as
\be
\psi \psi \sim 1 + \eta^{\tau_1/2}\sigma^{1-s_1}(1 + O(\sigma)) +\eta^{\tau_2/2}\sigma^{1-s_2}(1 + O(\sigma))  + \cdots
\ee
where $v=-\eta \sigma$, $u = 1/\sigma$. Let us focus on a particular power $\sigma^{1-s}$. Contributions of this form come from operators with spin $s' \geq s$, so in the lightcone limit $\eta \to 0$, the dominant contribution is the lowest-twist operator satisfying $s' \geq s$.  Denote by $\tau_s^*$ the twist of the spin-$s$ operator with smallest dimension. It was argued in \cite{Komargodski:2012ek} that $\tau_s^*$ is a monotonically increasing, convex function of spin, with $\tau_s^* \leq 2(d-2)$.\footnote{There, the argument held only above some unknown minimum spin, $s \geq s_{min}$. An identical argument can be made using the position-space sum rules, following the same logic. In this case we know that the sum rule is convergent for spin-2, so this establishes monotonicity and convexity for $s\geq 2$.} This guarantees that at the order $1/\sigma^{s-1}$, the leading contribution to the OPE indeed comes from $X$, which we defined to be the lowest dimension operator with spin $s$. 

If $\Delta_\psi < d-2$, an additional subtlety arises because there is an accumulation point in the twist spectrum at $\tau \sim 2 \Delta_\psi$ \cite{Fitzpatrick:2012yx,Komargodski:2012ek}. It is unclear whether the asymptotic lightcone OPE can be applied at orders in $\eta$ beyond the accumulation point. Therefore in what follows we assume the probe satisfies $\Delta_\psi > \tau_X/2$ (and that any other light scalars in the OPE that would lead to accumulation points are absent).

\subsection{Sum rule and positivity}

We can now simply repeat the argument of section \ref{s:properties}, inserting a factor of $\sigma^{s-2}$ into the sum rule integral to project onto the spin-$s$ contribution: $\oint d\sigma (1-G) \to \oint d\sigma \sigma^{s-2}(1-G)$.\footnote{Here we have followed \cite{Hofman:2016awc} to project onto a given spin. This method assumes that lower spin operators have integer dimensions, to avoid additional non-analytic contributions to the OPE which are subleading in $\sigma$ but leading in $\eta$, so can spoil the projection.  However the method can be generalized by subtracting these terms as well \cite{spinproj}.} The other steps are identical, leading to 
\be
-i\ c_{\psi\psi X} \langle  \overline{O(y=\delta)} \, \E_s \, O(y=\delta) \rangle =\frac{N_\delta}{\pi \lambda_X \eta^{\tau_s/2}}\lim_{R\rightarrow 0}\lim_{\eta\rightarrow 0}  \text{Re} \int_{-R}^R d\sigma\ \sigma^{s-2}(1-G(\sigma))\ 
\ee
for any $s\ge 2$ where $\lambda_X=\frac{2^{\Delta_X}   \Gamma \left(\frac{\Delta_X+s+1}{2}\right)}{ \sqrt{\pi}c_X  \Gamma \left(\frac{\Delta_X+s}{2}\right)}$. When $s$ is an even integer,  Rindler positivity ensures that the right hand side of the above sum rule is non-negative. Finally by acting with a rotation $R$ that rotates by $\frac{\pi}{2}$ in the Euclidean $y\tau$-plane, we can generalize (\ref{orotate}) for arbitrary spinning operators: 
\be\label{tpf}
\langle \overline{O(y=\delta)} \, \E_s \, O(y=\delta) \rangle= i^{s+1} \langle (R\sdot O)^\dagger(t=i\delta)\E_s (R\sdot O)(t=-i\delta)\rangle  \ .
\ee
This is derived in appendix \ref{app:rot3}.
Therefore, there is a constraint on the lowest dimensional operator at each even spin:
\be
(-1)^{\frac{s}{2}} c_{\psi\psi X} \E_s \ge 0\ .
\ee
Note that by taking $X \to -X$ it is always possible to choose $(-1)^{\frac{s}{2}} c_{\psi\psi X}>0$, and in that case we have a positivity condition similar to the ANEC. However, once this choice has been made for some coupling $\langle \psi \psi X\rangle$ we do not have the freedom to flip the sign for a different probe, $\langle \psi' \psi' X\rangle$.  This means that, like the stress tensor, the lowest-dimension operator of each spin must couple with the same sign to all possible probes. 

In fact these conclusions apply to the tower of operators appearing in any given $\psi\psi$ OPE. In theories with global symmetries, different probes may lead to different infinite families of constraints.

\subsection{Comparison to other constraints and spin-1-1-4 example}

In many cases, this sum rule implies the same sign constraints on CFT 3-point couplings that have already been deduced by other methods:  
\begin{itemize}
\item For $s=2$, the same results can be obtained from the ANEC using conformal collider methods \cite{Hofman:2008ar}. Therefore these constraints follow from monotonicity of relative entropy \cite{Faulkner:2016mzt}.
\item For $s>2$ with transverse polarizations $\varepsilon \sdot n = 0$, where $n$ is the null direction separating the wavepacket insertion from the insertion of $\E$, the results agree with deep inelastic scattering \cite{Komargodski:2016gci}. In examples where the results are available, these also agree with the lightcone bootstrap \cite{causality2,Hofman:2016awc}. If $O$ is a conserved current, then we can always choose a gauge where the polarizations are transverse.
\end{itemize}
For $s=2$ and $\varepsilon \sdot n \neq 0$ --- assuming $O$ is not a conserved current --- it was shown in \cite{Komargodski:2016gci} that the ANEC is stronger than deep inelastic scattering and the lightcone bootstrap.  Therefore, analogously at higher spin, we expect the condition $\E_s \geq 0$ to produce new constraints, stronger than any of these other methods, when $s>2$ and $O$ is not conserved.  

The simplest such case is $s=4$ with $O$ taken to be a spin-1, nonconserved operator $J$ of dimension $\Delta_J > d-1$. We will work out this example in detail and confirm that the sign constraints are indeed new.

The most general 3-point function $\langle JXJ \rangle$ consistent with conformal symmetry is written in appendix \ref{sec:3point}, following \cite{Costa:2011mg}.  It has four free numerical constants, $\alpha_{1,2,3,4}$. To derive the constraints, we apply the Hofman-Maldacena analysis to this 3-point function.  In practice, this amounts to computing the integral \eqref{finalhm}, with $O \to J$ and $T_{uu} \to X_{uuuu}$. Requiring this to be positive gives a constraint of the form
\begin{align}\label{positivity4}
{\bf \varepsilon}^\dagger {\bf A}  {\bf \varepsilon} \ge 0\ ,
\end{align}
where ${\bf \varepsilon} = (\varepsilon^+,\varepsilon^-, \vec{\varepsilon}^\bot)$ and ${\bf A}$ is a block diagonal matrix which depends on $\Delta_X$, $\Delta_J$ and the $\alpha_i$. The explicit formula can be found in \eqref{eq:141}. It follows that ${\bf A}$ must be a positive semi-definite matrix. Requiring the eigenvalues to be non-negative gives quadratic inequalities on the $\alpha_i$; the explicit form is unilluminating, so we will not write it explicitly, but it is easily found from \eqref{eq:141}.

Now let us compare to deep inelastic scattering \cite{Komargodski:2016gci}. Repeating their calculation for the present example, we find that the DIS constraints are
\begin{align}\label{ffdis}
&c_{\psi \psi X} (  2\alpha_4  -\alpha_1)\ge 0 \\
&c_{\psi \psi X} (\alpha_1 + (2+2\Delta_J-\Delta_X) \alpha_4 )\ge 0 \ .\nn
\end{align}
The constraints \eqref{ffdis} are identical to the constraint \eqref{positivity4} only if we set $\varepsilon^- = 0$. The constraints from the $\varepsilon^- \neq 0$ polarizations in \eqref{positivity4} are stronger, so these are new --- they do not follow from any of the known methods based on conformal collider bounds, the ANEC, relative entropy, DIS, or the lightcone bootstrap.  The special role played by $\varepsilon^-$ polarizations is analogous to the situation for the integrated null energy as described in \cite{Komargodski:2016gci}.

\vspace{1cm}
 
\bigskip

\textbf{Acknowledgments}
We are grateful to Nima Afkhami-Jeddi, John Cardy, Ven Chandrasekaran, Netta Engelhardt, Eanna Flanagan, Diego Hofman, Sachin Jain, Juan Maldacena, Samuel McCandlish, Eric Perlmutter, David Poland, Andy Strominger, John Stout, and Aron Wall for useful discussions.  The work of TH and AT is supported by DOE grant DE-SC0014123, and the work of SK is supported by NSF grant PHY-1316222. We also thank the Galileo Galilei Institute for Theoretical Physics and the organizers of the workshop Conformal Field Theories and Renormalization Group Flows in Dimensions $d>2$, as well as the Perimeter Institute for Theoretical Physics and the organizers of the It from Qubit workshop, which provided additional travel support and where some of this work was done.

\appendix

\section{Rotation of three-point functions}\label{app:rot3}
In this appendix, we will derive equation (\ref{tpf}). Let us start with the three-point function $\langle \overline{O(y=\delta)} \, \E_s \, O(y=\delta) \rangle$, where $O(y=\delta)$ is some arbitrary operator 
\be
O(y=\delta)=\sum_n c_n\ \varepsilon^n.{\cal O}^n(y=\delta)\ , \qquad \overline{O(y=\delta)}=\sum_n c_n^*\ \overline{\varepsilon}^n.{{\cal O}^n}^\dagger(y=-\delta)\ .
\ee
${\cal O}^n$'s are (not necessarily a primary) operators with any spin and $\E_s$ is defined in equation (\ref{defes}). $\overline{\varepsilon}$ is defined in the usual way $\overline{\varepsilon}^{\mu \nu ...}=(-1)^P(\varepsilon^{\mu \nu ...} )^*$, where $P$ is the total number of $t$ and $y$ indices. Let us now look at one of the terms of this three point function: $\langle \overline{\varepsilon}_1.{\cal O}_1^\dagger(-i \epsilon,-\delta)\ \E_s\ \varepsilon_2.{\cal O}_2(i\epsilon,\delta)\rangle$. Before performing the $u$-integral, this three point function, in general has the following branch cut structure:
\be
\includegraphics[width=0.4\textwidth]{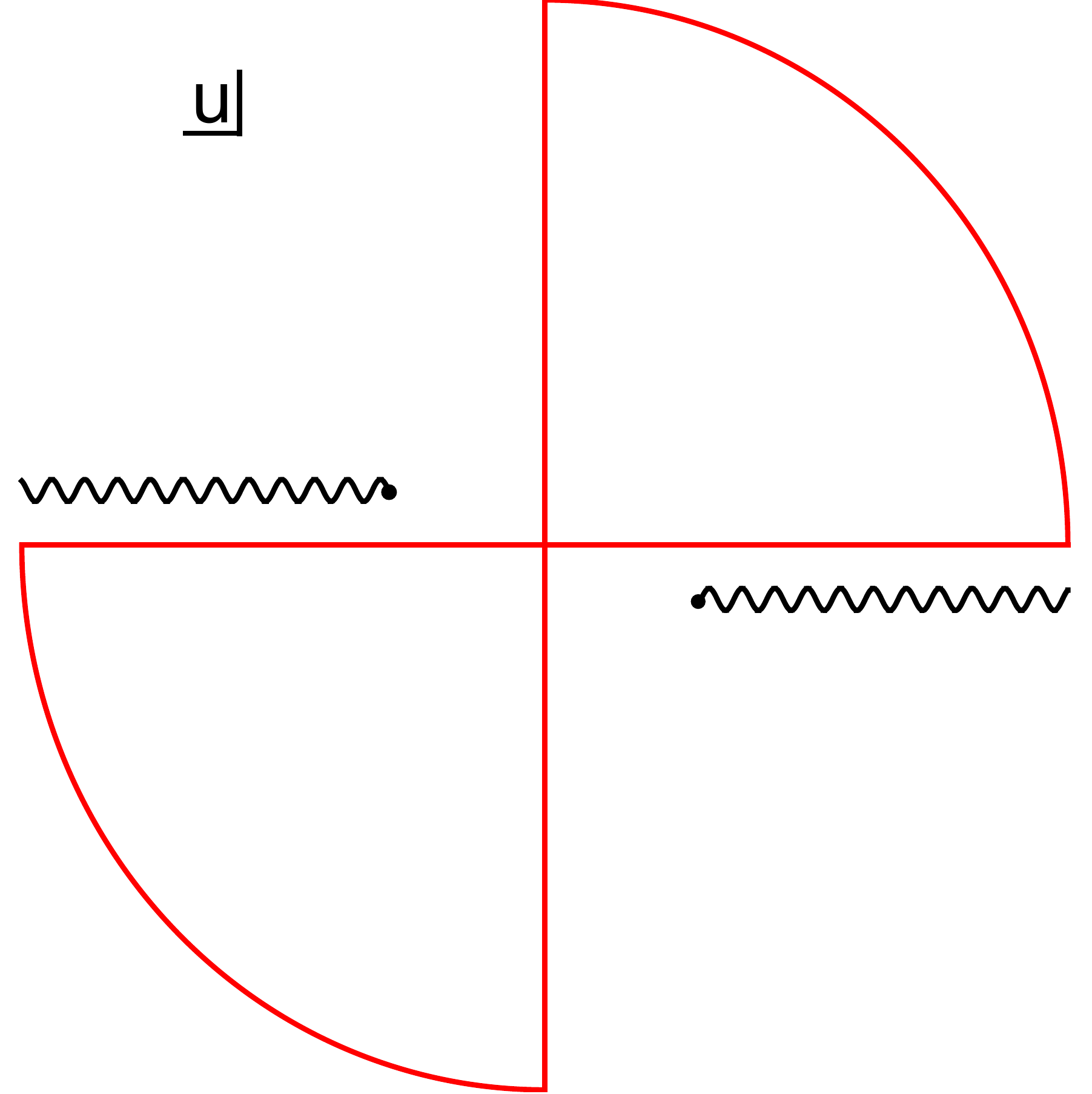}
\ee
Now, using the integration contour shown in red in the above figure, one can show that
\begin{align}
\langle \overline{\varepsilon}_1.{\cal O}_1^\dagger(-i \epsilon,-\delta)&\E_s \ \varepsilon_2.{\cal O}_2(i\epsilon,\delta)\rangle=i \langle \overline{\varepsilon}_1.{\cal O}_1^\dagger(0,-\delta)\int_{-\infty}^{\infty} du X_{uuu\cdots u}(i u)\ \varepsilon_2.{\cal O}_2(0,\delta)\rangle\ .
\end{align}
There is no ambiguity in the right hand side correlator and hence the $i\epsilon$ has been removed. Let us now look at the three point function $\langle \overline{\varepsilon}_1.{\cal O}_1^\dagger(0,-\delta) X_{uuu\cdots u}(i u)\ \varepsilon_2.{\cal O}_2(0,\delta)\rangle$.  This three-point function can be obtained by starting with some appropriate correlator in the Euclidean space: $(x_0, x_1,\vec{x})$ and performing an analytic continuation $x_0=i t, x_1=y$. On the other hand, we can start with the same Euclidean correlator and perform a different analytic continuation:  $x_1=i t, x_0=y$ to obtain a different Lorentzian correlator. Hence, these two Lorentzian correlators should be related to each other. More explicitly, one can show that 
\begin{align}
\langle \overline{\varepsilon}_1.{\cal O}_1^\dagger(y=-\delta)& X_{uuu\cdots u}(i u)\ \varepsilon_2.{\cal O}_2(y=\delta)\rangle\nonumber\\
&=i^s \langle \tilde{\varepsilon}^*_1.{\cal O}_1^\dagger(t=i\delta) X_{uuu\cdots u}(- u)\ \tilde{\varepsilon}_2.{\cal O}_2(t=-i\delta)\rangle\ ,
\end{align}
where,
\be
\tilde{\varepsilon}^{\mu \nu ...}=\left( \Lambda^\mu_\alpha \Lambda^\nu_\beta ...\right) \varepsilon^{\alpha \beta ...}\ , \qquad  \Lambda^\mu_\alpha =\begin{pmatrix}
  0 & -i & 0 \\
  -i & 0 & 0 \\
  0 & 0 & \mathds{1}
 \end{pmatrix}\ .
\ee
Therefore, finally we can write,
\be
\langle \overline{\varepsilon}_1.{\cal O}_1^\dagger(y=-\delta)\E_s \ \varepsilon_2.{\cal O}_2(y=\delta)\rangle=i^{s+1}\langle \tilde{\varepsilon}^*_1.{\cal O}_1^\dagger(t=i\delta) \E_s\ \tilde{\varepsilon}_2.{\cal O}_2(t=-i\delta)\rangle\ .
\ee

\section{Normalized three point function for $\langle J X J \rangle$}\label{sec:3point}

Here we write the matrix ${\bf A}$ and two point function of states we used in the paper explicitly. The three point function involving two same operators with spin one and another operator with spin 4 is given by

\begin{align}
\begin{split}
\langle J(x_1,\varepsilon_1)  X(x_2, \varepsilon_2) J(x_3, \varepsilon_3)\rangle& = \frac{1}{(x_{12}^2)^{\frac{\Delta_X}{2}+2}(x_{23}^2)^{\frac{\Delta_X}{2}+2}(x_{31}^2)^{\Delta_J+\frac{\Delta_X}{2}-1}} \{ \alpha_1 V_1 V_3 V_2^4  + \\
& \alpha_2 V_2^3 (V_3 H_{12} + H_{23} V_1) +\alpha_3 V_2^2 H_{12} H_{23}+  \alpha_4 V_2^4 H_{13} \}
\end{split}
\end{align} 

In which conformal building blocks are expressed by

\begin{align}
\begin{split}
&H_{ij} = -2 x_{ij}\cdot \varepsilon_j x_{ij}\cdot \varepsilon_i + x_{ij}^2 \varepsilon_i \cdot \varepsilon_j \\
&V_i \equiv V_{i,jk} =\frac{1}{x_{jk}^2} (x_{ij}^2 x_{ik} \cdot \varepsilon_i - x_{ik}^2 x_{ij}\cdot\varepsilon_i)
\end{split}
\end{align}
For state defined in \ref{lamo} expectation value of spin 4 operator X is given by
\begin{align}\label{eq:141}
\begin{split}
&c_{\psi \psi X}\langle \varepsilon^* \cdot  J| \int_\infty^\infty du  X_{uuuu}(0,u) | \varepsilon\cdot J \rangle  = \\
& (Vol) c_{\psi \psi X} \frac{2^{\Delta_X-4 \Delta_J -3}  \pi^{\frac{5}{2}}  \Gamma(2\Delta_J)  \Gamma(\frac{3+\Delta_X}{2})}{ \delta^{2\Delta_J}\Gamma(3+\Delta_J- \frac{\Delta_X}{2})\Gamma(3+\frac{\Delta_X}{2})\Gamma(1+\Delta_J+\frac{\Delta_X}{2}) } \times \\
&\{  |\varepsilon^+|^2 (4+\Delta_X) (2\Delta_J+ \Delta_X)(\Delta_X+ 2\Delta_J -2)(\Delta_X+2\Delta_J)(2 \alpha_4 -\alpha_1) \\
& +|\varepsilon^-|^2 (     4(4+2\Delta_J - \Delta_X)(12- \Delta_X^2 +2\Delta_J (4+\Delta_X)) \alpha_2 + \\
&2 (4+\Delta_X)(-2-2\Delta_J+\Delta_X) ( (8+4\Delta_J - 2\Delta_X) \alpha_3 + (\Delta_X - 2 \Delta_J) \alpha_4) \\
& - (96 + 16\Delta_J (4+\Delta_J) +4\Delta_J (1+\Delta_J) \Delta_X - 2(3+2\Delta_J) \Delta_X^2 + \Delta_X^3)\alpha_1
) + \\
&2 (\varepsilon^+ {\varepsilon^-}^* +\varepsilon^- {\varepsilon^+}^*) (2\Delta_J + \Delta_X) ((16+ 4 \Delta_X - \Delta_X^2 + 2 \Delta_J (2+\Delta_X)) \alpha_1\\
& - 2  (4+ 2 \Delta_J-\Delta_X) (2+ \Delta_X) \alpha_2 - 4(4+\Delta_X) \alpha_4) \\
&+ |\varepsilon^\bot|^2 (4+\Delta_X)(2\Delta_J+ \Delta_X)(\alpha_1 + (2+2\Delta_J- \Delta_X)) \alpha_4 \} 
\end{split}
\end{align}

By imposing unitarity $\Delta_J \ge d-1$ and the convexity condition for the twist of a spin-4 operator $ d-2 \le \Delta_X - 4 \le 2(d-2)$ \cite{Komargodski:2012ek} (in $d=4$), all gamma functions are positive. 

The volume term in three point function is canceled out by the same factor in the two point function, which is always positive:
\begin{align}
\begin{split}
&\langle \varepsilon^* \cdot J| \varepsilon\cdot J \rangle =\\
&(Vol) \;\frac{2^{3-2\Delta_J} c_J \Gamma(\Delta_1- \frac{3}{2}) }{ \delta^{2\Delta_J-3} \Gamma(1+\Delta_J)} \{  (|\varepsilon^+|^2 + |\varepsilon^-|^2) (2\Delta_J-4) - 2 (\varepsilon^+ {\varepsilon^-}^* +\varepsilon^- {\varepsilon^+}^*)  + |\varepsilon^\bot|^2 (\Delta_J -1) \}
\end{split}
\end{align}

\section{Free scalars}\label{s:free}
In this appendix, we show that the inequality $\E_s =  \int du X_{uuu\cdots u} \geq 0$ holds also for free scalar fields, with $s \geq 2$ an even spin. For $s=2$, this is the ANEC, proved for free scalars in \cite{Wald:1991xn,Klinkhammer:1991ki}. The OPE methods in the body of the paper do not immediately apply, because the expansion in twist has an infinite number of contributions already at leading order. Instead we will follow the derivation of the ANEC in \cite{Klinkhammer:1991ki}.

The all-null components of the conserved, symmetric, traceless spin-$s$ current in the theory of a free scalar takes the form
\be
X_{u\cdots u} = \sum_{i=0}^{s} a^s_i :\!(\p_u)^{i} \phi (\p_u)^{s-i}\phi\!:
\ee
where $a^s_i$ are known coefficients. (The explicit formula is equation (4.99) in \cite{Giombi:2009wh} but will not be needed.) Therefore after integration by parts, the generalization of the averaged null energy, up to an overall constant, is
\be\label{esddq}
\E_s \sim  \int du \, :\! (\p_u)^{s/2}\phi (\p_u)^{s/2}\phi\!: \ .
\ee
The overall coefficient does not matter, since we only need to show that $\E_s$ has a definite sign -- if it is non-positive, we can defined $X \to -X$ to make it non-negative. Classically, \eqref{esddq} is obviously sign-definite because the integrand is positive, but quantum mechanically this is true only after doing the integral. To proceed, we use the standard mode expansion for the scalar field operator $\phi = \int d^{d-1}\vec{k}[u_{\vec{k}} a_{\vec{k}} + h.c.]$ with $u_{\vec{k}} \sim e^{-ikx}$ to write the integrand of \eqref{esddq} as
\begin{align}
\int d^{d-1} &\vec{k} \int d^{d-1} \vec{k}'\big[ 2(-ik_u)^{s/2} (ik'_u)^{s/2} u_{\vec{k}} u^*_{\vec{k}'}  a^\dagger_{\vec{k}'}a_{\vec{k}} 
\\
& + (-ik_u)^{s/2}(-ik'_u)^{s/2} u_{\vec{k}} a_{\vec{k}} u_{\vec{k}'} a_{\vec{k}'} 
+ (ik_u)^{s/2}(ik'_u)^{s/2} u_{\vec{k}}^* a^\dagger_{\vec{k}} u^*_{\vec{k}'}a^\dagger_{\vec{k}'}
\big]\nn
\end{align}
The first term is obviously a non-negative operator, since it is $\int d^{d-1}\vec{k}u_{\vec{k}}a_{\vec{k}}$ times its Hermitian conjugate. The other terms can be negative (for example in squeezed states), but disappear upon integrating over the null ray in \eqref{esddq}, leaving only non-negative contributions. See \cite{Klinkhammer:1991ki} for details of these integrals, as well a careful demonstration that exchanging the order of the $u$-integral and $k$-integrals is justified in states with a finite number of particles and finite energy.

\end{spacing}

\end{document}